\documentclass[prb,twocolumn,showpacs,preprintnumbers,amsmath,amssymb,
               floatfix,superscriptaddress]{revtex4}

\usepackage{graphicx}
\usepackage{dcolumn}
\usepackage{bm}

\newcommand{\bra}[1]{\left\langle {#1} \right\vert}
\newcommand{\ket}[1]{\left\vert {#1} \right\rangle}
\renewcommand{\vec}[1]{\mathbf{#1}}

\begin{document}

\title{Signatures of nonadiabatic O$_2$ dissociation at Al(111):\\ First-principles fewest-switches study}%

\author{Christian Carbogno}%
\affiliation{%
Institut f\"ur Theoretische Chemie, Universit\"at Ulm, D-89069 Ulm, Germany
}%
\author{J\"org Behler}%
\affiliation{%
Lehrstuhl f\"ur Theoretische Chemie, Ruhr-Universit\"at Bochum, D-44780 Bochum, Germany
}%
\author{Karsten Reuter}%
\affiliation{%
Fritz-Haber-Institut der Max-Planck-Gesellschaft, Faradayweg 4-6, D-14195 Berlin, Germany
}%
\author{Axel Gro{\ss}}%
\affiliation{%
Institut f\"ur Theoretische Chemie, Universit\"at Ulm, D-89069 Ulm, Germany
}%

\date{\today}

\begin{abstract}
 Recently, spin selection rules have been invoked to explain the
 discrepancy between measured and calculated adsorption probabilities
 of molecular oxygen reacting with Al(111). In this work, we inspect
 the impact of nonadiabatic spin transitions on the dynamics of this
 system from first principles. For this purpose the motion on two distinct
 potential-energy surfaces associated to different spin
 configurations and possible transitions between them are inspected by
 means of the Fewest Switches algorithm. Within this framework we
 especially focus on the influence of such spin transitions on 
 observables accessible to molecular beam experiments.
 On this basis we suggest
 experimental setups that can validate the occurrence of such
 transitions and discuss their feasibility.
\end{abstract}

\maketitle

\section{Introduction}
Metal oxidation processes are of outstanding importance in various
industrial fields which range from combustion catalysis to
microelectronics. Accordingly, this research field has attracted considerable
scientific interest---both from experimentalists and theorists---over
the last decades. Although the most fundamental rules underlying such
processes were understood~\cite{Cabrera:1949p312} quite early, there
still exist some not yet elucidated phenomena, even for systems that
appear utterly simple at first sight. One of the most prominent and
intriguing examples is the oxidation of the (111) surface of
aluminum which exhibits various astonishing features. The exposure of
a clean aluminum surface to an oxygen atmosphere leads to the growth
of an oxide film, which then inhibits further corrosion. In spite of
the fact that the overall mechanism for the formation of this
Al$_2$O$_3$ stoichiometry has already been clarified by
qualitative~\cite{Gartland:1977p91} kinetic considerations, the
elementary key steps underlying this process are still topic of a
lively scientific debate. 

For instance, scanning tunneling microscopy (STM) images obtained
after exposing the clean aluminum surface to a minute dosage of
molecular oxygen show only single oxygen atoms in an average distance
of more than $80~\mbox{\AA}$ from each other~\cite{Brune:1992p222} on
the surface. Various hypotheses, e.g.,~the subsurface migration of
single oxygen atoms~\cite{Ciacchi:2004p544} or a cannonball like
abstraction dynamics~\cite{Neuburger:2000p3911,Binetti2000,Komrowski:2001p774} of
the impinging oxygen molecules, have been proposed to explain this
extraordinary finding. As a matter of fact, even the validity of the
original experiment has been questioned~\cite{SCHMID:2001p208}. The
contradictory nature of the proposed mechanisms, which are all backed
up by reasonable scientific findings, already shows how complicated
the experimental clarification of this issue is. 

However, at first sight the questions should be settled easily by
means of the current available simulation techniques~\cite{Gro:2002p4181}. 
Unfortunately, this is not the case for this particular system: Various studies carried out on the basis of
state-of-the-art adiabatic density functional
theory~(DFT)~\cite{Honkala:2000p49,Yourdshahyan:2002p38} were not even
able to qualitatively reproduce the low adsorption probability for
thermal $\mathrm{O}_2$ molecules at the clean aluminum surface
found in molecular beam experiments~\cite{Osterlund:1997p151}. If
even the simulation of such an elementary key step of the reaction
fails that dramatically, attempts to model the more complex features
of the dissociation dynamics within the framework provided by DFT are
foredoomed to fail, too.

Recently, however, DFT calculations addressing potential energy surfaces (PESs) of fixed 
spin states were able to reproduce the experimental data in a semiquantitative 
way. These studies suggest
that spin selection rules, which hinder~\cite{Behler:2005p19} an 
adiabatic spin transition from the initial $\mathrm{O}_2$ gas-phase 
triplet state to the singlet state of the adsorbed $\mathrm{O}$ atoms,
 are the rate determining factor in this process. For these investigations a
constrained density functional theory approach~\cite{Behler:2007p178}
was employed to compute the PESs of
$\mathrm{O}_2$ in different spin-configurations, and the lowered
adsorption probability was then correctly reproduced when
restricting~\cite{Behler:2008p322} the $\mathrm{O}_2$ molecules to move
on the spin-triplet potential-energy surface only. 

Although these studies yield a more than qualitative correspondence
with the experiment with respect to the initial sticking coefficient, 
they inherently incorporate an approximation that
is  ultimately not satisfactory. By restricting the motion 
to the triplet PES alone, the impinging $\mathrm{O}_2$ molecules are
not able to relax to the correct final singlet state of the adsorbed
atoms while dissociating. Therefore we extend these
previous studies by considering the dissociation dynamics on multiple
spin potential-energy surfaces, allowing transitions between them
within the mixed quantum-classical \textit{Fewest
 Switches}~\cite{Tully:1990p207} algorithm as proposed by Tully,
which has been successfully employed in the modeling of
various~\cite{Bach:2005p335,Carbogno:2007p338} molecule-surface
processes before. As detailed below, we can thereby reproduce the
measured sticking coefficient and thus reconfirm the earlier more restricted 
modeling. Additionally, our simulations allow to study the occurrence,
the nature and the influence of the spin transitions in detail. 
On the basis of these calculations, we were already
able to propose~\cite{Carbogno:2008p187} measurements that can clearly
prove the nonadiabaticity of the studied reaction. In this work we
further study such potential experimental setups and especially focus
on their feasibility. Therefore, we explicitly inspect how various
uncertainties underlying the theoretical modeling, as for instance 
the frozen substrate approximation, affect the practical realization
of such scattering experiments in the lab.

The proposed measurements can verify the fundamental assumption behind 
our approach that
hindered spin transitions are indeed the rate determining factor in
this process. This is of particular importance due to the fact that
speculations~\cite{Mosch:2008p829,Livshits:2009p3797} have arisen
which do not relate the low sticking coefficient of thermal oxygen
molecules to the nonadiabatic dynamics. Rather, shortcomings of the
state-of-the-art but still approximative exchange-correlation GGA
functional employed in the DFT calculations are held responsible for
the discrepancy found between the adiabatic simulations and the
experiment. Such conjectures are hard to confute theoretically, at
least as long as calculations with an improved exchange-correlation 
functional are numerical too costly to be
performed accurately on a large scale. Certainly, this dictates also
further research along these lines. However, the herein proposed
experiments can clarify this severe doubt already now and thus
establish a solid founding for further research on the complex
phenomena associated with the oxidation of aluminum surfaces.

In this paper we will first review the \textit{ab initio} methods
underlying the simulations in Sec.~\ref{Theo}. Since the DFT
techniques used for the calculation of the total energy of the system
as well as the techniques employed to interpolate this set of data to
a smooth potential-energy surface have already been discussed in
detail~\cite{Behler:2007p178,Behler:2007p324} before, we will focus on
the method used to calculate the dynamics on multiple potential-energy
surfaces, i.e.,~Tully's Fewest Switches~\cite{Tully:1990p207}
algorithm. Especially, we will discuss the nature of the simulated
electronic states and the nonadiabatic coupling between them. In
Sec.~\ref{RES} we will then present the results of such simulations,
whereby a special focus lies on the occurrence and on the role of the
nonadiabatic transitions. In this context we will also present
possible experimental approaches able to identify the presence of
the spin transitions, whereby we will also discuss the
feasibility of the proposed experiments. We
especially focus on the uncertainties underlying the employed
theoretical approach and try to quantify them by additionally
inspecting the influence of different coupling strengths and the role
of the surface mobility.

\section{Theoretical methods}
\label{Theo}

\subsection{Mixed Quantum-Classical Dynamics}

When inspecting the dynamics of a system that includes nonadiabatic
transitions, the electronic and nuclear degrees of freedom cannot be
tackled independently and subsequently, as done in the
Born-Oppenheimer approximation. Unfortunately, a full quantum
dynamical description of both the electronic and nuclear degrees of
freedom is computationally far too demanding for the present system. 
However, for the nuclei
usually a classical treatment is sufficient.  Still, the fast
electronic and slow nuclear degrees of freedom have to be inspected
simultaneously, whereby a mutual self-consistent feedback between them
is crucial~\cite{Tully:1998p169} to achieve a correct description of
the system. In this work we employ a mixed quantum-classical method,
namely the \textit{Fewest Switches Surface Hopping} algorithm as
proposed by Tully~\cite{Tully:1990p207}, to achieve this goal. For a
concise summary of this approach, the total Hamiltonian for a system
including electronic ($\mathbf{r}$) and nuclear ($\mathbf{R}$) degrees
of freedom
\begin{eqnarray}
H & = & T_{\mathbf{R}} + T_{\mathbf{r}}  + V_{\mathbf{r}} +
V_{\mathbf{r}\mathbf{R}}+  V_{\mathbf{R}}\label{eq:H} \nonumber\\
 & = & T_{\mathbf{R}} + H_{el}(\mathbf{r},\mathbf{R})+ V_{\mathbf{R}}
\label{Hzusammen}
\end{eqnarray}
will serve as a starting point. In Eq.~(\ref{Hzusammen}), $T$ denotes
the kinetic energy operators, $V$ the respective electrostatic
potentials and all terms depending on the electronic coordinates are
subsumed in $H_{el}$. The total wave function for this system can then
be expressed as
\begin{equation}
\Psi(\mathbf{r},\mathbf{R},t) = \sum_i c_i(t)\Phi_i(\mathbf{r},\mathbf{R}) \ ,
\label{eq:expansion}
\end{equation}
whereby the basis functions~$\Phi_i(\mathbf{r},\mathbf{R})$ used in
the expansion, also referred to as electronic states in this work, are
not necessarily eigenfunctions of the electronic
Hamiltonian~$H_{el}$. 
For each of these basis functions, we can derive a
Hamilton-Jacobi equation for the nuclear dynamics, in which the potential
\begin{equation}
V_{ii}(\vec{R}) = \bra{\Phi_i(\mathbf{r},\mathbf{R})} H_{el}(\mathbf{r},\mathbf{R}) + V_{\mathbf{R}}\ket{\Phi_i(\mathbf{r},\mathbf{R})}
\label{HJ}
\end{equation}
resulting from the integration
over the electronic degrees of freedom indicated by the bra-ket
notation, is specific to the thereby employed basis
function~$\Phi_i(\mathbf{r},\mathbf{R})$. 
Consequently, an equation of
motion for each particular electronic state included in the
expansion~(\ref{eq:expansion}) can be derived. Whenever the nuclear
masses of interest are significantly larger than the one of hydrogen,
so that quantum effects in the dynamics of the slow degrees of freedom
can be safely neglected~\cite{Gro:1997p238}, the motion can be described
by classical equations of motion
\begin{equation}
M\frac{d^2}{dt^2}\mathbf{R} =-\nabla_{\mathbf{R}} V_{ii}(\mathbf{R}).
\label{eq:classical}
\end{equation}
Still, these equations do just describe the motion on one single PES
but not the transitions between different PESs.

If we insert the full expansion~(\ref{eq:expansion}) into the
time-dependent Schr\"odinger equation associated with the
Hamiltonian~$H$, we gain a set of coupled differential
equations~\cite{Tully:1990p207,Bach:2005p335}
\begin{equation}
i\hbar \dot{c}_k = \sum_j c_j(t)\left(V_{kj}(\mathbf{R}(t))-i\hbar
\mathbf{\dot{R}}(t)\cdot\mathbf{d}_{ij}(\mathbf{R}(t))\right) ,
\label{eq:coeff}
\end{equation}
which describe how the expansion coefficients related to the
individual electronic states evolve along a
trajectory~$\mathbf{R}(t)$. Beside the non-diagonal potential matrix
elements~$V_{kj}(\mathbf{R}(t))$, also the \emph{nonadiabatic coupling
 vector}
\begin{equation}
\mathbf{d}_{ij}(\mathbf{R})=
\left \langle {\Phi_i(\mathbf{r},\mathbf{R})} \right \vert
\nabla_{\mathbf{R}}\left \vert{\Phi_j(\mathbf{r},\mathbf{R})} \right \rangle,
\label{Kopplungsvektor}
\end{equation}
which designates the dependence of the chosen basis
functions~$\Phi_i(\mathbf{r},\mathbf{R})$ on the nuclear positions,
enters the equations. By integrating Eq.~(\ref{eq:coeff}) along a
trajectory, the changes in the occupation of the different electronic
states during the dynamics, i.e.,~the time evolution of the diagonal
elements of the density matrix $a_{ij}(t)=c_{i}(t)^*c_{j}(t)$, can be
inspected for a specific trajectory.  

This fact can be exploited to combine
Eq.~(\ref{eq:classical}) and (\ref{eq:coeff}) to a surface hopping
method: At each time step~$\Delta t$, the system is in exactly
\textbf{one} electronic state $k$ that determines which potential
$V_{kk}$ is used in Eq.~(\ref{eq:classical}) to compute the classical
trajectory.  On the other hand, the complete set of equations
(\ref{eq:coeff}) is integrated along the trajectory to determine the
occupation in \textbf{all} electronic states of the expansion. If by
any chance another electronic state from the one that actually
determines the nuclear dynamics becomes dominant, the trajectory
``switches'' to this state: From then, this new state will
determine the potential in the nuclear equation of motion, as long as
no other switch occurs. To actually decide whether and when such a
``surface hop'' occurs, a Metropolis~\cite{Metropolis:1953p854} like
algorithm is combined with the switching probability proposed by
Tully~\cite{Tully:1990p207}
\begin{equation}
P_{k\rightarrow j}(t) = \frac{\dot{a}_{jj}(t)}{a_{kk}}\Delta t \ .
\end{equation}
This so called \textit{Fewest Switches} approach, which distinguishes
this method from comparable surface hopping approaches~(see
Ref.~[\onlinecite{Tully:1998p169}] and references therein), fulfills
two essential conditions. For a large number of computed trajectories,
the population in each state statistically matches the probabilities
given by the density matrix $a_{ii}$. Thereby, this distribution is
achieved with as few surface hops as possible. Please note that this
algorithm allows switches to occur at any point along the trajectory
$\mathbf{R}(t)$, even if the potential energies differ (e.g.,~$\Delta
V = V_{jj}(\mathbf{R})- V_{kk}(\mathbf{R})\neq 0$). For such surface
hops with $\Delta V \neq 0$, special care has to be taken with respect
to energy conservation. If the kinetic energy of the nuclei is smaller
than the potential difference ($T_{\mathbf{R}} < \Delta V$), the
switches are rejected. Conversely, if the potential energy difference
is larger than the kinetic energy ($T_{\mathbf{R}} > \Delta V$), the
velocities have to be rescaled to fulfill $T_{\mathbf{R}}^j -
T_{\mathbf{R}}^ k = -\Delta V$. This constraint is not unique because
the direction of the velocity adjustment can be chosen
freely. Considerations from semi-classical theory suggest to rescale
the velocities along the nonadiabatic coupling vector, since the
forces associated with nonadiabatic transitions
typically~\cite{Herman:1984p650,Coker:1995p654} point in its
direction.

Although the herein given description of the Fewest Switches method is
adequate to understand the calculations presented below, it is
certainly not exhausting. The interested reader is referred to the
original publications describing this method~\cite{Tully:1990p207,
 HeadGordon:1995p726, Tully:1998p169}.

\subsection{The inspected electronic states}
\label{InElSt}

\begin{figure*}
       \begin{minipage}{0.475\linewidth}
       \includegraphics[clip,width=\linewidth]{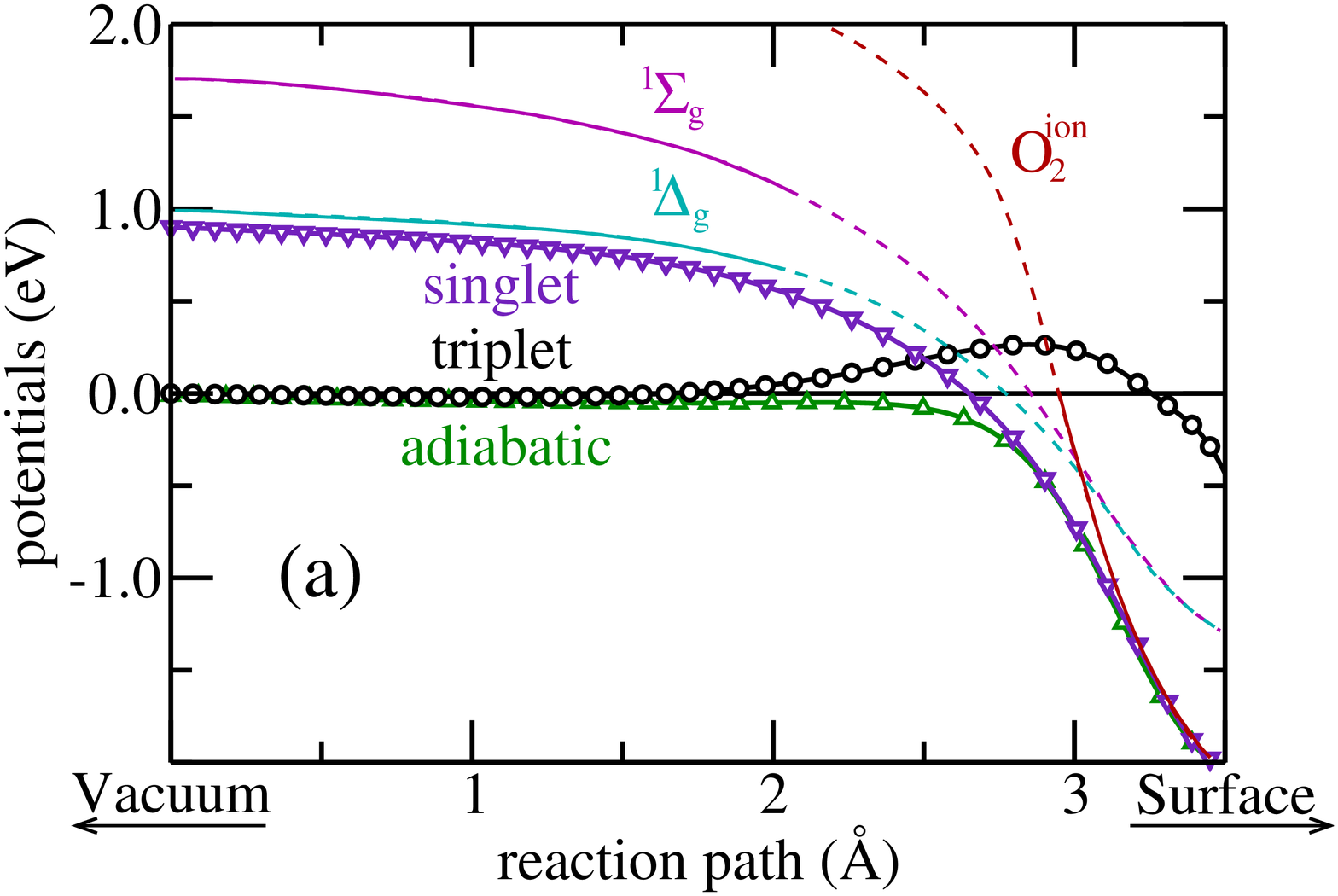}
       \end{minipage}
       \hfill
       \begin{minipage}{0.475\linewidth}
       \includegraphics[clip,width=\linewidth]{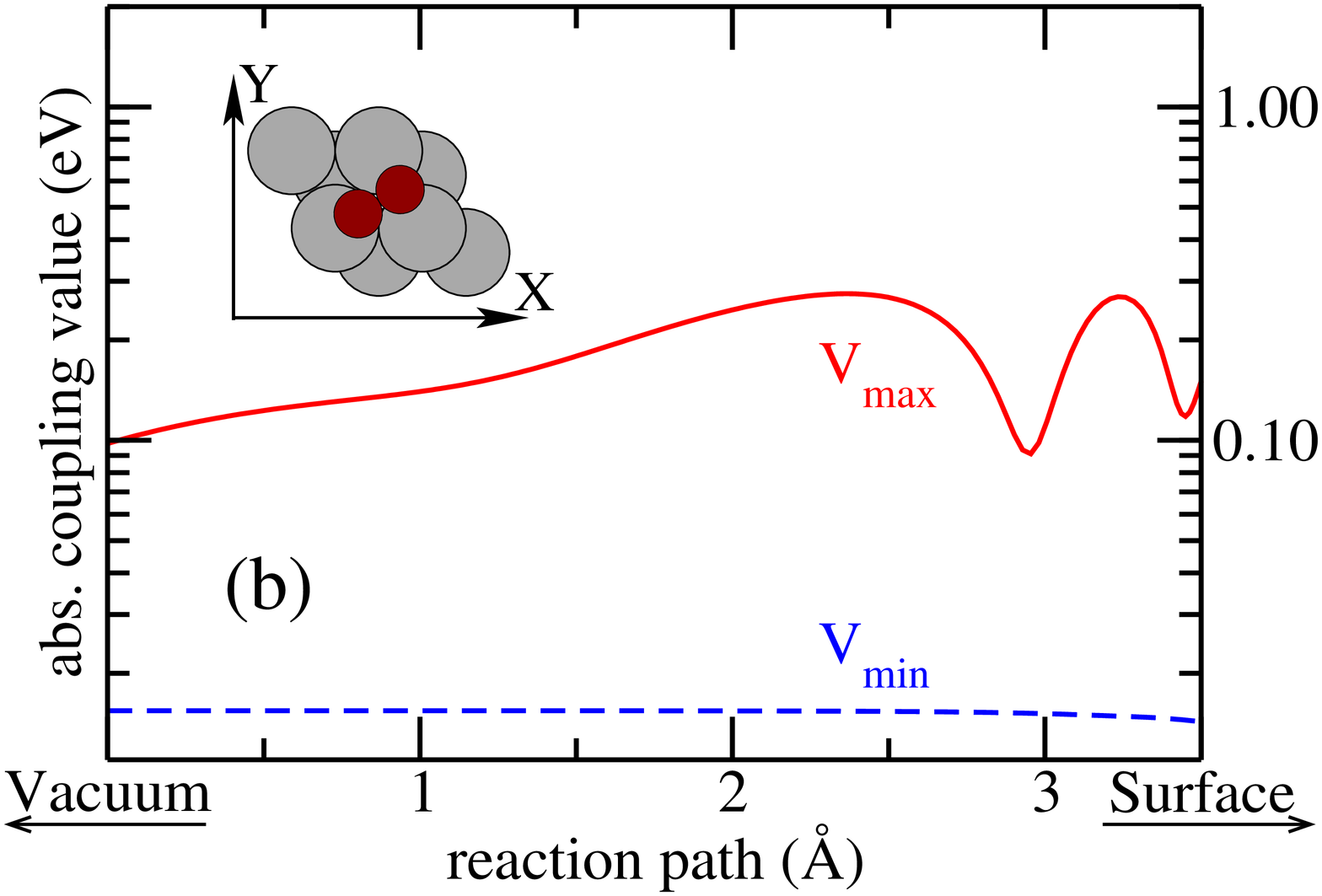}
       \end{minipage}
       \caption{Schematic plot of the potential-energy surfaces along
         the minimum energy path in the triplet state for
         dissociation in parallel alignment over the fcc site: Beside
         the adiabatic (spin-polarized), the triplet (constrained
         spin-polarized) and the singlet (adiabatic non-spin-polarized)
         PESs, which have been determined by means of DFT, also the 
         expected functional
         form of the potentials associated with the first two excited
         singlet states and the so-called ``ionic'' state are shown
         in Fig.~$(a)$. In Fig.~$(b)$ the lower and the upper bound
         derived for the coupling of the triplet and the singlet PES
         are plotted along the minimum energy path of Fig.~$(a)$.}
\label{SingletScheme4}
\end{figure*}

The first essential step in the simulation of nonadiabatic reactions
with the surface hopping method is the identification of the
electronic states of interest, i.e.,~the basis used for the
expansion~(\ref{eq:expansion}). Since a distinct
PES~$V_{ii}(\vec{R})$, which has usually to be calculated individually
under huge numerical effort, is associated to each one of these
electronic states, it is desirable to achieve a correct description of
the system with as few expansion terms as possible. Fortunately, the
inclusion of the adiabatic ground state and one additional excited
state is typically already sufficient to model nonadiabatic processes
accurately~\cite{Tully:1998p169}. In our case an analysis of the
adiabatic calculation---a so called \textit{diabatization}---serves as
a starting point for the determination of the electronic states that
are relevant in the inspected system's dynamics, since such a
\textit{diabatic}\footnote{We use the adjective ``diabatic'' for
 electronic states whose wave function does not depend on the nuclear
 coordinates~\cite{Tully:1990p207}. In the quantum chemistry
 community such an approach is often known as ``crude
 diabatic''~\cite{Mead:1982p246}.} representation allows for a more
natural and intuitive interpretation of the occurring electronic
transitions.

As shown for an exemplary geometry in Fig.~\ref{SingletScheme4}, the
PES associated with the initial state of the reaction, i.e.,~the
triplet ground state of the isolated oxygen molecule, resembles the
adiabatic ground state PES closely at large molecule-surface distances
and is thus the first state included in the surface hopping
approach. Not too surprisingly, this triplet PES does not coincide
with the adiabatic PES in the limit of small oxygen-surface distances:
In this final state of the reaction, the dissociated oxygen atoms adsorbed 
on the surface exhibit a singlet type ground state electronic structure, 
whereby an additional charge transfer~\cite{Honkala:2000p49,Behler:2007p178} 
from the surface onto the molecule occurs. Although there might be 
multiple charged states of the oxygen molecule involved~\cite{Katz2004} 
in the actual electron transfer process, we can qualitatively incorporate 
such effects in this schematic model by one effective PES. This so called 
ionic PES resembles the adiabatic potential closely in the late part of 
the dynamics and approaches the lowest lying charge-transfer state in the 
vacuum limit. Its respective energy gap can be estimated from the clean 
aluminum surfaceÕs work function and the isolated oxygen moleculeÕs electron 
affinity to be 4 eV with respect to the triplet ground state.
In between these two  states,
two additional singlet states of the neutral oxygen molecule,
i.e.,~the $^1\Delta_g$ and the $^1\Sigma_g^+$ state, are
located. Although these two states do neither correspond to the
initial nor to the final state of the reaction, they play a crucial
role in the dynamics. In the proximity of the barriers present on the
triplet PES they definitely are as relevant as the respective triplet
and ionic PES, since they exhibit comparable curve crossings with the
triplet state.

It would certainly be desirable to include all of these excited states
in the simulation of the nonadiabatic dynamics. However, even the
first principles calculation of one single PES alone is utterly costly in terms of
computing power. But even more importantly, the $^1\Delta_g$ and the
$^1\Sigma_g^+$ state exhibit a strongly correlated, multi-determinant
wave function~\cite{Behler:2007p178,Behler:2008p322,Slipchenko:2002p2} that is not described correctly
by single-determinant theories such as Hartree-Fock and DFT. As a
matter of fact, the current exchange-correlation functionals, which do
not depend on the symmetry of the electronic structure, do not allow
to discriminate~\cite{Gunnarsson:1980p266,VonBarth:1979p265} between
the $^1\Delta_g$ and the $^1\Sigma_g^+$ state at all. This becomes
evident already when inspecting the \textbf{isolated} oxygen molecule
by an adiabatic non-spin-polarized DFT calculation, which must yield a
singlet state by construction. In this approach, however, the
respective wave functions of the two singlet states
mix~\cite{CarbognoFuture} in equal parts, so that the
resulting triplet-singlet gap lies between the correct values that are
associated with the $^1\Delta_g$ and the $^1\Sigma_g^+$ state,
respectively. When additionally including the aluminum slab in such an
adiabatic non-spin-polarized DFT calculation, the actual triplet-singlet
gap is lowered due to a spurious and unphysical charge transfer from
the surface that occurs even at large molecule-surface separations:
The current GGA exchange-correlation functionals erroneously favor
such a charge separation due to their incorrect convexity with respect
to fractional charges~\cite{Cohen:2008p5}. As a consequence the
thereby computed PES mimics the first excited $^1\Delta_g$ singlet
state at large distances from the surface, as shown in
Fig.~\ref{SingletScheme4}. In the proximity of the surface, this PES
coincides with the one associated with the adiabatic ground state,
since the charge transfer between oxygen and aluminum is not
restrained in the underlying adiabatic non-spin-polarized DFT
calculations. Thus such an approach enables us to effectively describe
both the charge transfer effects close to the surface as well as the
influence of the neutral singlet states within \textbf{one} single
PES. However, such an approximative but still adequate description of
the manifold of excited states also has some drawbacks. For the
calculation of the coupling elements we have to disentangle the
contributions from the various electronic states once again, as
discussed in the next section.

For the sake of readability the adiabatic spin-polarized PES will henceforth 
be referred to as ``adiabatic'', the constrained spin-polarized triplet PES
as ``triplet'' and the adiabatic non-spin-polarized PES as ``singlet''. All 
three potential energy surfaces, which explicitly depend on the six degrees 
of freedom of the oxygen molecule, have been constructed 
in the frozen substrate approximation by interpolating large sets of   
\textit{ab initio} data. A detailed discussion of the underlying constrained 
and unconstrained DFT simulations~\cite{Behler:2007p178,Behler:2008p322}, 
of the neural network interpolation~\cite{Behler:2007p324} and of the dynamics
on a single PES~\cite{Behler:2005p19,Behler:2008p322} can be found in the cited publications.

\subsection{The Coupling}
\label{Coupl}
For the derivation of the coupling elements we will again rely on the
diabatic representation~(see Sec.~\ref{InElSt}) of the triplet and the
singlet state~($\ket{\Psi_t},\ket{\Psi_s}$), for which the
nonadiabatic coupling vector
\begin{equation}
\mathbf{d}_{ts}(\mathbf{R})=
\bra{\Psi_t}\nabla_{\mathbf{R}}\ket{\Psi_s} = 0
\label{Kopplungsvektor_TS}
\end{equation}
vanishes exactly at each point in space due to the trivially fulfilled
orthogonality of the respective spinors. Consequently, the evolution
of the density matrix in Eq.~(\ref{eq:coeff})---and hence the
electronic transitions---is completely determined by the non-diagonal
elements of the electronic Hamiltonian
\begin{equation}
V_{ts} = \bra{\Psi_t} \mathcal{H}_e \ket{\Psi_s}.
\end{equation}
To disentangle the various interactions contributing to this coupling,
we decompose the singlet wave function, which is actually a mixture of
three distinct states~(see Sec.~\ref{InElSt}), into its individual
components:
\begin{equation}
\Psi_{s} = A \cdot \left( \frac{1}{\sqrt{2}}\left(\Psi_{\Delta} + \Psi_{\Sigma}\right)\right) + B\cdot \Psi_{\mbox{\tiny ion}},
\label{WavefunctionSinglet}
\end{equation}
i.e.,~the first excited $^1\Delta_g$ singlet state~$\Psi_{\Delta}$, the second
excited $^1\Sigma_g^+$ singlet state~$\Psi_{\Sigma}$ and the ``ionic'' state~$\Psi_{\mbox{\tiny ion}}$,
which adiabatically incorporates all charge transfer effects and is identical to the singlet ground state 
close to the surface. Thereby the complex parameters~$A$ and $B$ are determined by the adiabatic mixing of the neutral and the ionic state described before.\\

In a first step, we neglect the contributions arising from the coupling to the ionic state and focus on the isolated molecule in the vacuum limit, where no charge transfer is supposed to happen. Under these assumptions the coupling can be  simplified to:
\begin{equation}
V_{ts} = \frac{1}{\sqrt{2}}\left(\bra{\Psi_t} \mathcal{H}_e \ket{\Psi_{\Delta}} + \bra{\Psi_t} \mathcal{H}_e \ket{\Psi_{\Sigma}}\right) .
\label{TSQC}
\end{equation}
Since the spinors  present on the left and on the right hand side of the Hamiltonians in Eq.~(\ref{TSQC}) are orthogonal to each other, the only interaction that does not vanish in the evaluation of this matrix element is the spin-orbit coupling~(SOC)
\begin{equation}
V_{ts} =  \frac{1}{\sqrt{2}}\left(\bra{\Psi_t} V_{\mbox{\tiny SOC}} \ket{\Psi_{\Delta}} + \bra{\Psi_t} V_{\mbox{\tiny SOC}} \ket{\Psi_{\Sigma}}\right) .
\label{TSQC2}
\end{equation}
The first of the two coupling terms in Eq.~(\ref{TSQC2}), i.e.,~the SO
coupling of the triplet ground state to the first excited
$^1\Delta_g$~singlet state, vanishes in first order approximation,
since selection rules prohibit a direct mixing of the two
states~\cite{Kearns:1971p11,Schweitzer:2003p190}. Solely, an indirect
coupling~\cite{Minaev:1980p123,Klotz:1986p231} through the highly
excited $^3\Pi_g$~state, which is at least two orders of magnitude
smaller than the direct SO coupling of the triplet ground state to the
$^1\Sigma_g^+$~state discussed below, is possible. Thus the
overwhelming portion of the spin-orbit coupling is provided by the
second term, i.e.,~the SO coupling of the triplet ground state to the
second excited $^1\Sigma_g^+$~singlet state. For the isolated oxygen
molecule, this non-diagonal matrix element has been calculated as a
function of the oxygen-oxygen distance with the SOC
module~\cite{JANSEN:1999p41} provided by the quantum chemical
simulation package \textit{Molpro}~\cite{MOLPRO} on the basis of wave
functions determined\footnote{The V5Z correlation consistent basis set
 and Molpro's default numerical integration parameters have been used
 for the simulations. The convergence with respect to both the basis
 set and these parameters have been carefully~\cite{CarbognoFuture}
 inspected.} by \textit{Multireference Configuration
 Interaction}~\cite{Werner:1988p139} before. The thereby determined
coupling $V_{ts}$---as well as the derived lifetime of $11.16$~s for
the $^1\Sigma_g^+$ state---is in excellent agreement with experimental
and theoretical data found in literature~\cite{Kearns:1971p11,
 Schweitzer:2003p190, Minaev:1996p80}. With respect to the
oxygen-oxygen separation, which is the only degree of freedom the
coupling actually depends on in these calculations, it is a linear,
slowly decreasing~(almost constant) function, even up to huge
separations~(see Fig.~\ref{SingletScheme4}). If actually a factual
dissociation of the isolated molecule does not affect this coupling,
also the interaction with the aluminum, which exhibits only a small
SOC~\cite{Schwarz:2004p133} due to its low mass, will not alter it
significantly. Consequently,
\begin{equation}
V_{\mbox{\tiny min}} = \frac{1}{\sqrt{2}}\bra{\Psi_t} V_{\mbox{\tiny SOC}} \ket{\Psi_{\Sigma}},
\label{VMIN}
\end{equation}
the absolute value of which is plotted in Fig.~\ref{SingletScheme4},
can be regarded as a lower bound for the true
coupling present in the system.

On the basis of this minimal coupling determined above and the
PESs described in Sec.~\ref{InElSt} we can construct the diabatic
potential matrix
\begin{equation}
\mathbb{H} = 
\begin{pmatrix}
\bra{\Psi_t} \mathcal{H}_e \ket{\Psi_t} &  \bra{\Psi_t} \mathcal{H}_e \ket{\Psi_s}\\     
\bra{\Psi_s} \mathcal{H}_e \ket{\Psi_t} &  \bra{\Psi_s} \mathcal{H}_e \ket{\Psi_s}
\end{pmatrix}
=
\begin{pmatrix}
E_t   &  V_{\mbox{\tiny min}}\\     
V_{\mbox{\tiny min}}^* &  E_s\\     
\end{pmatrix}
\end{equation}
and diagonalize it to obtain the respective adiabatic representation of the potentials
\begin{equation}
E_{\mbox{\tiny adia}} = \frac{1}{2}\left(E_s + E_t \pm
\sqrt{
\left(E_s-E_t\right)^{2}+4 \left\vert V_{\mbox{\tiny min}}\right\vert}\;\right)
\label{Diagonalization}
\end{equation}
for this minimal coupling. Due to its minute value, however, only a
small deviation from the diabatic potentials occurs at all, as already 
evident from the
minute gap of approximately $2 V_{\mbox{\tiny min}}\approx 30$~meV at
the avoided crossing point. This fact also
highlights the main shortcoming of this minimal coupling element: The
minute splitting at the avoided crossing is not able to recover the
adiabatic ground state potential that has been determined by the
unconstrained DFT calculations before. However, this fact also allows
to turn the tide: By forcing the left hand side of
Eq.~(\ref{Diagonalization}) to correspond to the adiabatic potential
determined by DFT before, we can invert the diagonalization to obtain
a complementary value for the coupling element
\begin{equation}
\left\vert V_{\mbox{\tiny max}} \right| = \sqrt{E_s E_t-E_s E_{\mbox{\tiny adia}}^{\mbox{\tiny DFT}}-E_t E_{\mbox{\tiny adia}}^{\mbox{\tiny DFT}}+\left(E_{\mbox{\tiny adia}}^{\mbox{\tiny DFT}}\right)^2}. 
\label{CouplAdia}
\end{equation}
In contrast to the one-dimensional minimal coupling, this coupling
term~$\left\vert V_{\mbox{\tiny max}} \right|$ is a true
six-dimensional function of the oxygen coordinates, since the inverse
diagonalization~(\ref{CouplAdia}) can be evaluated at each point in
space for which respective triplet, singlet and adiabatic potential
values are available. Accordingly, this coupling also depends on the
respective rotational and center of mass degrees of freedom so that it
does not exhibit such a constant behavior along the minimum energy
path as the SOC~(see Fig.~\ref{SingletScheme4}). Due to the spurious
charge transfer occurring in the adiabatic calculations, however, this
maximal coupling does not vanish in the vacuum limit as one would
expect. Additionally, the inverse diagonalization procedure itself is
generally expected to overestimate the coupling~\cite{Wu:2006p99}, so
that $ V_{\mbox{\tiny max}}$ can be regarded as an upper bound for
the true coupling present in the system. In fact we believe
that the maximal coupling is closer to the true coupling strength than the minimal coupling.

\begin{figure*}[t]
       \begin{minipage}{0.475\linewidth}               
       \includegraphics[clip,width=\linewidth]{Fig2a.eps}
       \end{minipage}
       \hfill
       \begin{minipage}{0.475\linewidth}
       \includegraphics[clip,width=\linewidth]{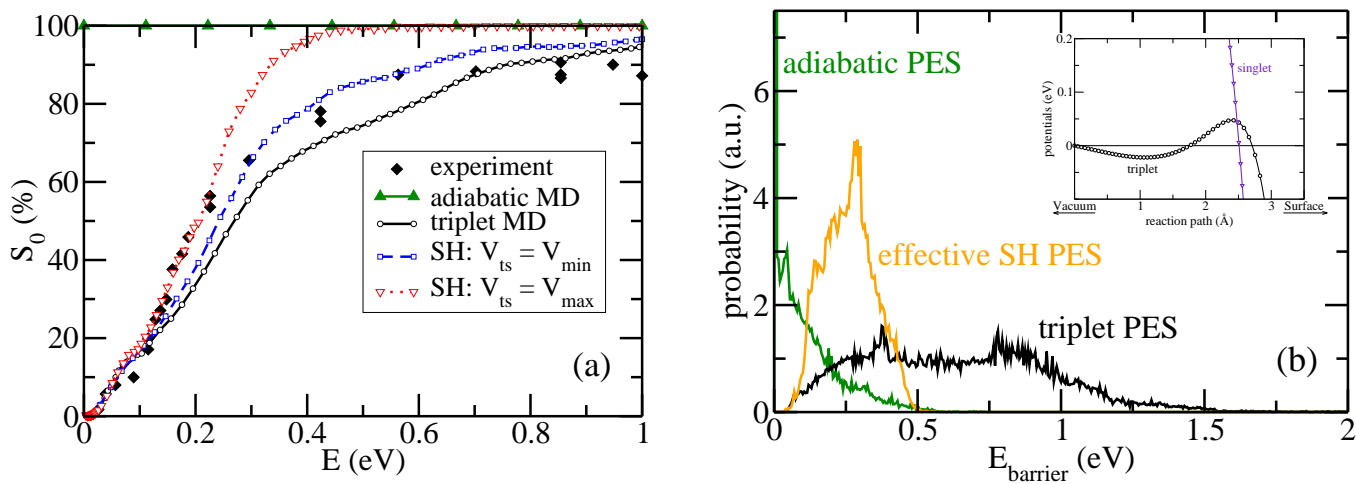}
       \end{minipage}
       \caption{$(a)$ Sticking coefficient~$S_0(E)$ as
         computed by molecular dynamics~(MD) on the triplet
         potential-energy surface and by Tully's surface hopping
         method~(SH) for both a maximal and a minimal coupling of the
         triplet and the singlet PES. The
         respective experimental data~\cite{Osterlund:1997p151} is
         shown as well. $(b)$ Area-normalized
         histograms of the barriers found on the individual PESs by
         the two-dimensional Nudged Elastic Band technique (see text). The
         potentials along the minimum energy path over the bridge
         site are shown in the inset.}
\label{StickCoeff}
\end{figure*}

In the case of such a maximal coupling, a diagonalization of the
respective diabatic Hamiltonian matrix yields a ground state potential
that resembles the adiabatic PES exactly, as required by
construction. Consequently all barriers previously present in the
diabatic picture vanish as well, so that surface hopping simulations
of thermal molecules, for which the diabatic calculations yield a
sticking probability of $2$\%~(see Sec.~\ref{StickCoeff2}), lead to a
constant sticking coefficient of $100$\% in the adiabatic
picture. This alarming discrepancy between the two representations can
be traced back to our approximative description of the manifold of
excited states by a single effective PES. In the limit of a large coupling our
model describes the two-level system consisting of the triplet and the
ionic state, but does not account for the presence of the other two
neutral singlet states.

In the diabatic picture this is indeed a
justified approximation, since due to the minute coupling the two
singlet states just provide for a negligible amount of additional
nonadiabatic transitions with respect to the ionic state. 
This is no longer true in the adiabatic picture: In this
representation, the potential-energy surfaces of states that are
strongly coupled are also strongly distorted at avoided curve
crossings.  The smaller the coupling of the respective states is in
the diabatic picture, however, the more probable transitions between
the adiabatic PES become. By not including the singlet states and
their minute coupling in the adiabatic representation of our two-level
model, we are de facto ignoring the adiabatic PES that matters most,
i.e.,~the one to which a transition will almost certainly occur at the
avoided crossing. To avoid these problems without the cumbersome
construction of another PES~(see Sec.~\ref{InElSt}), we performed all
the calculations presented here in the diabatic representation. The
price to pay is that the effective barriers in the simulations are
slightly overestimated, since nuclear tunneling effects and
resonances, which would be inherently
incorporated~\cite{Tully:1998p169} in an adiabatic picture, are
neglected in such a diabatic representation. Fortunately, these
effects are typically negligible in high-dimensional, heavy
systems~\cite{Bach:2000p340,Bach:2001p341} such as the one addressed
in this work.

\subsection{Calculational Details}
In all presented MD and SH simulations the initial conditions 
are chosen to model a molecule in the vacuum limit, in which it
does not interact with the aluminum surface. Therefore the initial center-of-mass 
distance between the oxygen molecule and
the surface is always~$5$~\AA.
Also, a certain amount of energy~$E$ is
assigned to the translational motion perpendicular to the surface.
The initial oxygen-oxygen distance and vibration is chosen to model the zero point
energy of the free molecule~($\approx 0.1$~eV) in the so called quasiclassical
approximation~\cite{Gro:1997p238}. 
In the beginning no kinetic energy is assigned to the rotational and to the 
lateral center-of-mass degrees of freedom, but their initial values are also
chosen randomly to uniformly examine all possible initial conditions.
To sample the whole phase space and to converge the calculations with respect
to the electronic transitions at least $2,000$ trajectories have been computed 
for each of the data points in the following figures. When a higher accuracy 
was needed, as in Sec.~\ref{HiNRG} for instance, up to $2,000,000$ trajectories
have been computed. In each case the calculation of the individual trajectories is 
stopped when one of the following conditions is reached: 
Molecules are regarded as dissociated, if the oxygen-oxygen
distance exceeds~$2.0$~\AA~($\approx 160$\% of the equilibrium bond length of
the free molecule), and as reflected, if the center-of-mass distance to the surface 
exceeds~$5$~\AA. The maximal time interval for one trajectory is chosen so
that virtually all molecules eventually fulfill one of these two conditions.

Naturally, the convergence of the results with respect to the presented calculational details 
and with respect to the numerical integration routines has been examined extensively~\cite{CarbognoPHD}.

\section{Results and discussion}
\label{RES}

\subsection{The Sticking Coefficient}
\label{StickCoeff2}

The sticking coefficient~$S_0(E)$ at normal incidence, i.e.,~the
probability of dissociative adsorption as a function of the incident
translational energy~$E$ for a molecular beam oriented perpendicular
to the surface, is the basic observable that we are going to discuss
while investigating the nonadiabatic dynamics of this system. As shown
in Fig.~\ref{StickCoeff}, it exhibits an S-like shape in the
experiments---a typical evidence for an activated process. MD
calculations on the adiabatic PES alone, however, yield a sticking
coefficient of constantly 100\% even at thermal incident
energies~\cite{Behler:2005p19} due the almost complete absence of
barriers~\cite{Yourdshahyan:2001p82,Behler:2007p178} on the respective
PES. Conversely, MD simulations on the triplet PES reproduce
the measurements more than satisfactorily, as already discussed in the
introduction.

With respect to the MD simulations performed on the triplet PES
alone~\cite{Behler:2005p19,Behler:2008p322}, the inclusion of nonadiabatic transitions within the surface
hopping method changes the simulated sticking coefficient only
quantitatively, but not qualitatively.  At large incidence energies
there is obviously no substantial difference at all: The impinging
oxygen is able to overcome the barriers anyway, regardless of any
occurring electronic transitions. In contrast to this, molecules at
medium velocities are not always able to overcome the barriers on the
triplet PES. Whereas such molecules are repelled in the triplet MD
calculations, they can dissociate by switching to the singlet PES in
the SH simulations, as soon as they reach the respective crossing
point~(see Fig.~\ref{SingletScheme4}a). The probability for such an
electronic transition to occur depends on the coupling of the two
electronic states so that the relative increase in dissociation found
in the SH simulations varies with the coupling strength. Last but
not least, the triplet MD and the SH simulations yield the same
sticking coefficient at the smallest incident energies below
$150$~meV: Molecules that are not even capable of reaching the
crossing point are also not able to benefit from the possibility of an
electronic transition. This might seem surprising at first, since one
may expect at least some triplet-singlet crossing points to occur
also in this low energy regime. However, this is not the case: As
exemplarily shown for the bridge site in the inset of
Fig.~\ref{StickCoeff}b, the crossing point and the top of the barrier
almost coincide even in geometries that exhibit barriers well below
$150$~meV. As a consequence, the sticking coefficient is not
affected by the switches in these setups, in spite of the fact that there is a high probability for such transitions at these low velocities.

These effects can be studied more systematically on the basis of the
actual barrier heights that the molecules experience on the individual
PESs in the sudden approximation. To determine this statistical
distribution we inspected the minimum energy path on a large number of
elbow plots by a two-dimensional, self-implemented\footnote{Whereas
 the oxygen-oxygen and the oxygen-surface distance were optimized
 during the calculations, the other degrees of freedom, i.e.~the
 rotational ones and the lateral surface coordinates, were kept
 fixed. A total of 350,000 of such configurations were inspected
 for each of the histograms shown in Fig.~\ref{StickCoeff}b. In each
 run 20~images were used and the location of the dissociated
 final state, if there was any in the respective elbow plot, was
 determined by a crude bisection method. The initial image was kept
 fixed in the vacuum at $5.0\mbox{ \AA}$ distance from the surface at
 the equilibrium bond-length of an isolated oxygen molecule.}
\textit{Climbing-Image Nudged Elastic
 Band}~\cite{Henkelman:2000p228,Henkelman:2000p186} method. Both the
histograms of the barriers found on the adiabatic PES and on the
triplet PES are shown in Fig.~\ref{StickCoeff}b. The effective
barriers experienced by the impinging molecules in the surface hopping
simulations, however, depend on the probability for an electronic
transition and thus both on the coupling and on the velocity of the
molecules. At fast velocities and for a minute coupling, the molecules
stay on the triplet PES and thus behave as in the triplet MD
simulations. In this limit, the barrier distribution of the triplet
state describes the dynamics best. Conversely, the impinging molecules
switch to the singlet state almost instantaneously while bypassing the
crossing seam at low velocities and for huge couplings. In this limit,
the energetic location of the crossing points can be regarded as an
\textit{effective barrier} distribution. 

As shown in Fig.~\ref{StickCoeff}b, the average height of these effective
barriers has also been
determined\footnote{To construct a single and smooth PES on which the
 crossing point indeed are the highest barriers, we inspected the
 adiabatic representation of the triplet-singlet system for a
 vanishing coupling ($\ll 1$~meV).} by means of the
\textit{Climbing Image Nudged Elastic Band} method. The resulting
barrier distributions substantiate the
previous findings of the simulations: The adiabatic PES is unactivated
for the large majority of geometries and even in the case of a
hindered dissociation the arising barriers are typically smaller than
$0.5$~eV. In contrast to this, the triplet PES exhibits exclusively
activated entrance channels with a broad distribution of barriers that
ranges from approximately $40$~meV up to more than $1.5$~eV. These
barriers are drastically reduced by the occurrence of electronic
transitions at the crossing points, the energetic height of which is
$0.5$~eV at most. However, the lowest occurring barriers are not
modified due to the reasons discussed before.

Please note that the great majority of these barriers lies in the
entrance channel, i.e.,~before the oxygen molecule starts to dissociate and
the oxygen atoms move apart. Accordingly, no distinct influence of the
vibrational state on the dissociation probability is
expected. Nevertheless, the slight vibrational enhancement found in the
experiment~\cite{Osterlund:1997p151} at small incident energies can be
reproduced~\cite{CarbognoPHD} by the simulations and can be traced
back to an effective lowering of the barriers due to an adiabatic
energy transfer~\cite{Gro:1996p895} occurring in the nuclear dynamics
from the vibrational degree of freedom onto the translational one.

\subsection{High Energy Scattering}
\label{HiNRG}

\begin{figure}
\centering
\includegraphics[clip,width=.9\linewidth]{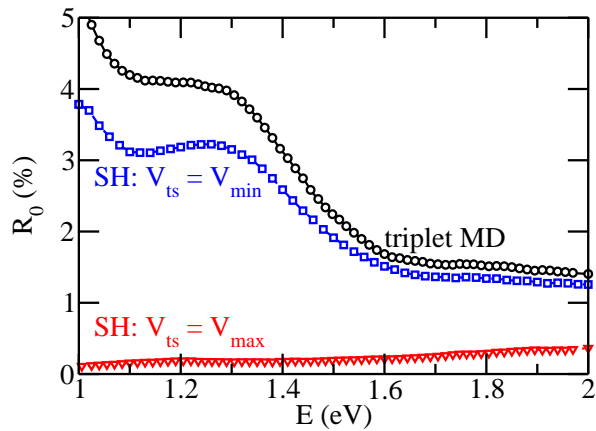}
\caption{Reflection coefficient at high incident translational energies as computed by MD on the triplet PES and by Tully's surface hopping method~(SH) for a minimal and a maximal coupling of the triplet and the singlet PES.}
\label{HighNRG_6D} 
\end{figure}

Although the hitherto presented simulations reproduce the experimental
data successfully, this does not explicitly prove the hypothesis~\cite{Behler:2005p19} that
nonadiabatic spin-transitions indeed play a decisive role in the
dynamics of the triplet oxygen molecules at the aluminum surface. 
Therefore we searched for a possible
experimental setup that is able to unambiguously demonstrate the occurrence of such
transitions. The fact that the triplet PES exhibits barriers as high
as $1.5$~eV might indeed offer a route towards experimental
verification: For incident energies that are larger than the
triplet-singlet gap ($\approx 1$~eV), energy conservation does no
longer prohibit a conversion of triplet into singlet molecules during
the backscattering. If such singlet molecules were detected in the
reflection channel, the nonadiabaticity of the scattering process
could be experimentally proved. 

To investigate the feasibility of such a measurement, we focus on the
reflection coefficient~$R_0(E) = 1 - S_0(E)$, i.e.,~the probability
for a molecule with perpendicular incidence to be repelled back into
the vacuum as a function of the initial translational energy~$E$. As
shown in Fig.~\ref{HighNRG_6D}, we indeed find a small amount of
molecules ($<5$\%) that are reflected in the triplet MD simulations
due to the barriers on the triplet PES. The larger the incident
energies, the less probable such backscattering events obviously
become. For incident energies above $1.5$~eV, for which the triplet
PES does not exhibit any barriers anymore, the reflection coefficient
does however not drop to zero but levels off at a value of
$1.4$~\%. In this energy regime, the translational movement is so fast
that the oxygen molecules approaching the surface in an unfavorable
geometry are not able to dissociate in the small time interval before
being backscattered, since they do not follow the minimum
energy path anymore. 

Not too surprisingly, the SH simulations with a minimal coupling
closely resemble the triplet MD calculations. For such huge velocities
and a minute coupling, oxygen behaves diabatically, i.e.,~it just
stays on the triplet PES in almost all trajectories. In contrast to this, in the
simulations with maximal coupling a significant amount of molecules still relaxes to the
singlet PES, which exhibits no barriers, as soon as the crossing point
is passed. Accordingly, the
associated SH simulations yield only a minute reflection coefficient
that slightly \textbf{increases} with larger incident energy. Even for such a huge coupling, the probability for electronic
transitions decreases with larger velocities, so that for large
incident energies more and more molecules stay on the more repulsive
triplet PES and are backscattered. Still, this effect is almost
negligible, even for the fastest molecules. Regardless of the
coupling, the already small reflection probabilities are even further
diminished as soon as one gives up the frozen substrate
approximation~(see Sec.~\ref{frozsub}). Even worse, less than
$3$\% of these reflected molecules are found in the singlet state upon
reflection, so that only $0.01$\% of all incident molecules result in
a backscattered singlet molecule at most, even in in the frozen
substrate approximation. This fact makes an experimental detection of
the nonadiabatic transition in the presented way factually impossible.

\begin{figure}
\centering
\includegraphics[width=0.9\linewidth,clip]{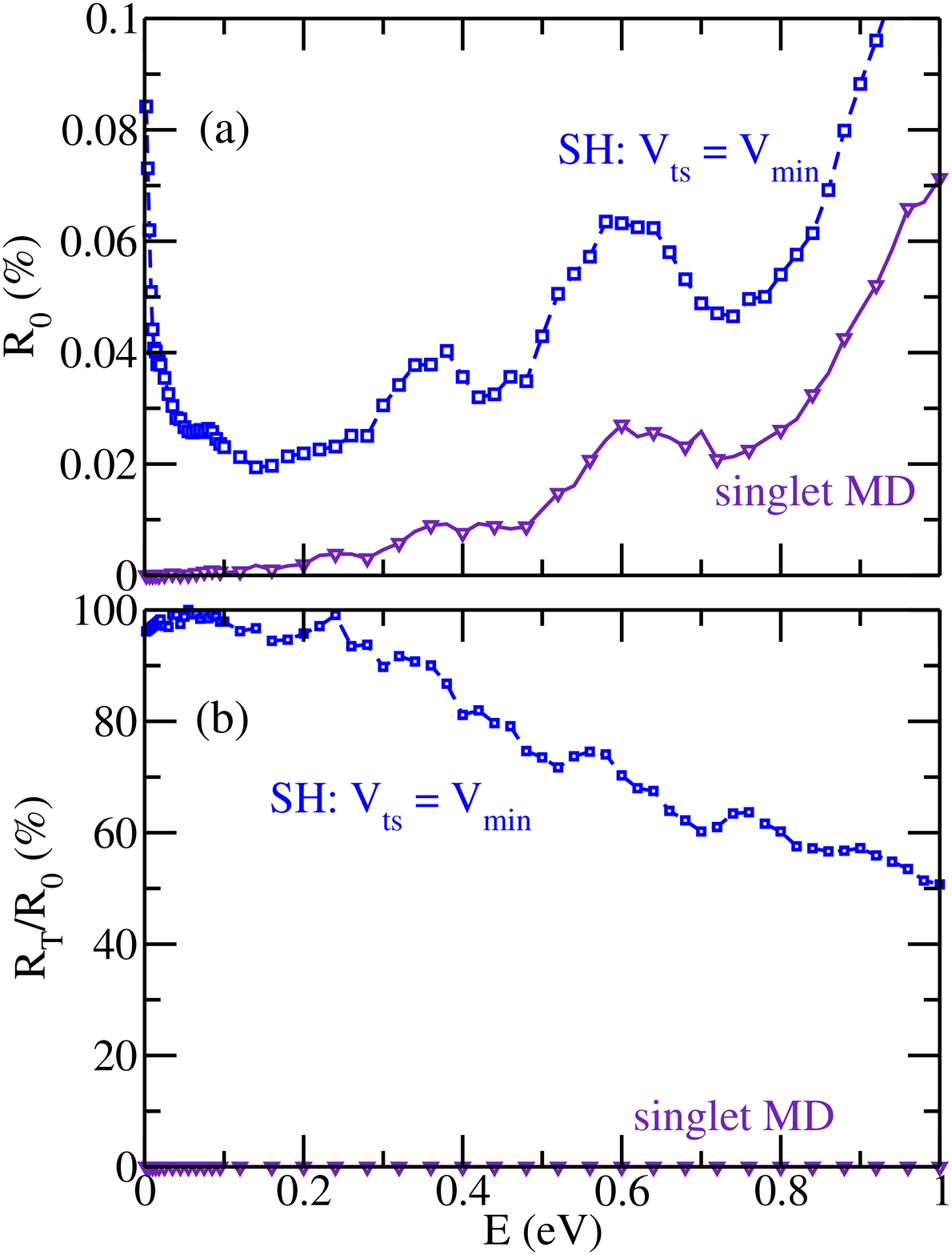}
\caption{$(a)$ Reflection coefficient~$R_0$ as function of the incident kinetic 
energy of singlet oxygen molecules, as calculated by the SH method with a 
minimal coupling. Additionally, the respective data obtained by MD simulations 
on the singlet PES alone is shown. $(b)$ Relative triplet yield~$R_T/R_0$ 
for both calculations.}
\label{SingletReflMin}
\end{figure}

\subsection{Singlet Scattering}
\label{Singl}
The sticking coefficient~$S_0(E)$ of triplet molecules discussed above
does not provide an unambiguous experimental evidence for the occurrence of
nonadiabatic spin-transitions. This is principally caused by the fact
that this typically inspected observable is not particularly sensitive
to the occurring electronic transitions. For the same reason
also a pure triplet MD approach yields a more than satisfactory
agreement with the experimental data~(see Sec.~\ref{StickCoeff2}). To
actually nail down the influence of the spin-transitions on the
reaction dynamics experimentally, a complementary process that is
significantly more sensitive to the transitions has to be
found. Indeed, we could identify~\cite{Carbogno:2008p187} the
dissociative adsorption of singlet oxygen molecules to provide for
experimental observables that strongly depend on the occurrence of
such spin-transitions. A comparable strategy has already been used in
a recent combined experimental and theoretical
study~\cite{Burgert:2008p157} to demonstrate the importance of spin
selection rules for O$_2$ interacting with small Al anion clusters
($\sim$10 to 20 atoms). In this work, an odd/even pattern in the
reactivity of triplet oxygen has been found as a function of the number
$n$ of atoms in the Al clusters, whereas no such pattern has been
found for the reaction of singlet molecules with these clusters. These
studies motivated further studies addressing the dynamics of singlet
oxygen, for example in the combination with other metal surfaces such as
Ag(100)~\cite{Alducin:2008p263}.

In our simulations we studied the dynamics of singlet oxygen molecules
impinging in normal incidence on the aluminum surface at various
kinetic energies. In the following, we will concentrate on two complementary
observables that will allow us to understand the nonadiabatic
processes active in the reaction. Firstly, we will inspect the
\textbf{total} amount of backscattered molecules as function of the
incident translational energy, i.e.,~the \textit{reflection
 coefficient}~$R_0(E)$. In a second step, we will also inspect the
\textbf{relative} amount of molecules that are backscattered in the
triplet state, i.e,~the \textit{relative triplet
 yield}~$R_T(E)/R_0(E)$, which turns out as the actual fingerprint of the
nonadiabatic transition. 

As shown in Fig.~\ref{SingletReflMin}, the probability of
backscattering rises with increasing kinetic energy in the case of a
minimal coupling: Although the singlet PES does not exhibit any
barriers at all, some molecules at high velocities are reflected on
this PES before the oxygen molecules are able to dissociate. The
higher the incident energy, the more probable such a process obviously
becomes. Therefore the respective SH simulations resemble MD
simulations on the singlet PES closely, apart from a constant
offset. This difference is caused by the fact that in the backscattering
process a reallocation of the translational energy onto the other
degrees of freedom occurs. Such vibrational or rotational excited
molecules might not have the translational energy to escape the
surface on the singlet PES anymore. While such molecules become
trapped close to the surface and then finally dissociate in the
singlet MD calculations, such molecules can escape in the SH
simulations by switching to the triplet PES thus gaining the energy of
the singlet-triplet splitting which helps them to leave the
surface. This mechanism leads to an exponentially increasing
reflection coefficient at small incident energies, for which almost
all backscattered molecules leave the surface in the triplet state.

\begin{figure}
\centering
\includegraphics[width=0.9\linewidth,clip]{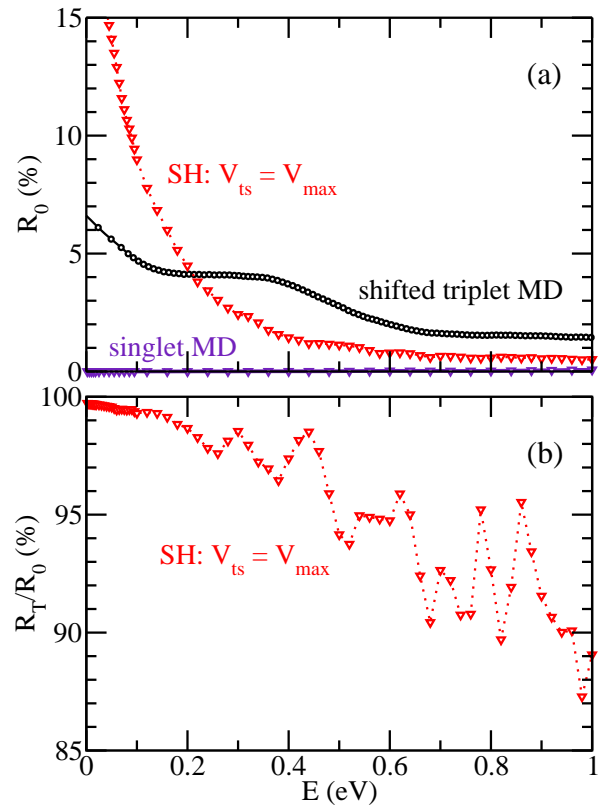}
\caption{Same as Fig.~\ref{SingletReflMin}, but for maximal coupling.}
\label{SingletReflMax}
\end{figure}

\begin{figure*}[t]
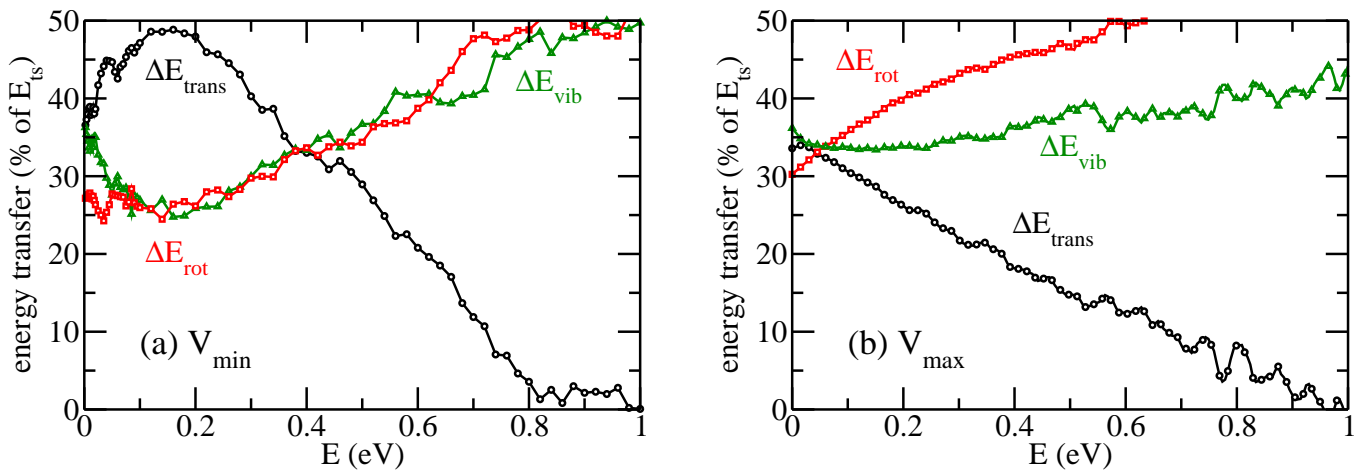

\centering
\begin{minipage}[t]{1.0\linewidth}
\begin{minipage}[t]{0.475\linewidth}
\includegraphics[width=1.0\linewidth,clip]{Fig6a.eps}
\end{minipage}
\hfill
\begin{minipage}[t]{0.475\linewidth}
\includegraphics[width=1.0\linewidth,clip]{Fig6b.eps}
\end{minipage}
\end{minipage}
\caption{Average kinetic energy gained by singlet molecules when being converted to triplet molecules in the backscattering process at the aluminum surface 
for a $(a)$ minimal coupling; $(b)$ maximal coupling.}
\label{Etrans}
\end{figure*}

Intuitively one might expect the same mechanism to be active also in
the case of the maximal coupling~(see Fig.~\ref{SingletReflMax}). This
is indeed the case at large incident energies, for which a small
amount of molecules is backscattered on the singlet PES and then
partially escapes the surface in the triplet state. The increased
coupling favors the transition to the triplet state and thus increases
both the total amount of reflected molecules as well as the relative
fraction of triplet oxygen. A totally different mechanism, which is
not active in the simulations with the minimal coupling, causes an
exponential increase of the reflection coefficient for low and medium
incident energies in the simulations with the maximal coupling. When
approaching the surface, the strong coupling destabilizes the singlet
state, so that a decay to the triplet ground state becomes extremely
probable even while the crossing seam is approached. Thus the great
majority of the molecules hit the surface on the triplet PES, on which
there are barriers as high as $1.5$~eV present~(see
Fig.~\ref{StickCoeff}b), so that a backscattering event is orders of
magnitude more likely than on the unactivated singlet
PES. Consequently, one would expect the singlet reflection
probability to be similar to the one of triplet molecules that hit the
surface with a comparable kinetic energy~$E+\Delta E_{\mbox{\tiny
   ts}}$.

A comparison with such a shifted triplet MD reflection
coefficient surprisingly reveals that the respective reflection
probability of the singlet molecules is even higher, as shown in
Fig.~\ref{SingletReflMax}. During the nonadiabatic relaxation to the
triplet state, which does not exclusively occur at the crossing point
but also in its proximity where the two PESs do not coincide, the
surplus of potential energy is reallocated onto all degrees of
freedom. Since this allotment is performed along the nonadiabatic
coupling vector (see Sec.~\ref{Theo}), not all the potential energy
stored in the triplet-singlet gap~$E_{\mbox{\tiny ts}}$ is transferred
onto the translational degrees of freedom, which are principally
responsible for surmounting the barriers. Rather, also the
vibrational, rotational and lateral center-of-mass degrees of freedom,
which do not promote the dissociation as strongly or might even hinder
it, gain a certain portion of kinetic energy, which is then missing in the
perpendicular translational motion. Consequently, these simulations
are not directly comparable to the triplet MD, in which the complete
amount $E+\Delta E_{\mbox{\tiny ts}}$ is assigned to the translational
degree of freedom perpendicular to the surface.

The fact that there are two different mechanisms active at low
incident energies depending on whether the minimal or the maximal
coupling is employed becomes evident when inspecting the velocities
that the triplet molecules exhibit in the reflection
channel. Fig.~\ref{Etrans} shows the average kinetic energy gain that
an incident singlet molecule experiences when being backscattered as a
triplet. Since this data has been derived from the velocities that the
reflected triplet molecules exhibit back in vacuum, the plots include
both the energy rearrangement due to the nonadiabatic transition, but
also all adiabatic energy redistribution processes that occur during the
backscattering dynamics. The two different mechanisms active at low
incident energies result in a qualitatively different reallocation of
the energy. At an incident energy of $0.2$~eV for instance, the
strongest energy gain occurs in the translational degrees of freedom
for the minimal coupling, whereas it occurs in the rotational channel
for the maximal coupling.

The fact that the triplet-singlet conversion results in a kinetic
energy gain also facilitates the realization of the proposed
experiment. At low incident energies, there is no need to detect the
electronic character of the backscattered molecules by,
e.g.,~phosphorescence, since energy conservation allows to disentangle
the contributions of the two molecular species by measuring the
kinetic energy distribution of the reflected
molecules~\cite{KrasnovskyJr:1993p963}, even in the case of a mixed
triplet/singlet incident molecular beam. If any of the degrees of
freedom exhibits an energy that is not compatible with the
translational energy that the molecular beam initially exhibits, a
transition from the singlet state must have happened. It is thus
sufficient to perform a TOF measurement to unambiguously detect a
significant amount of triplet oxygen through the respective kinetic
energy gain.

\subsection{The frozen substrate approximation}
\label{frozsub}

\begin{figure*}[t]
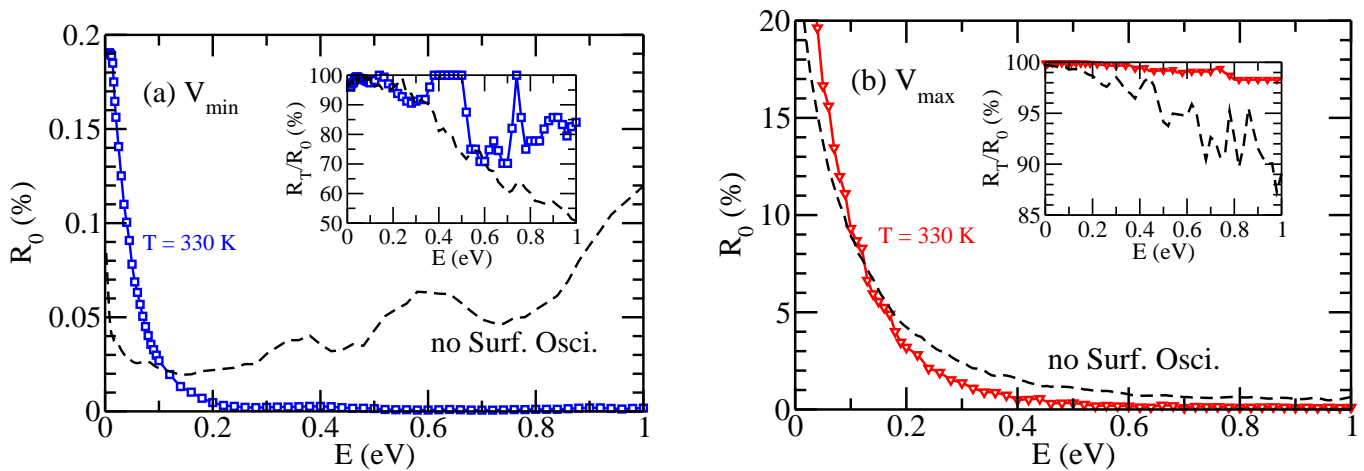

\centering
\begin{minipage}[t]{1.0\linewidth}
\begin{minipage}[t]{0.475\linewidth}
\includegraphics[width=1.0\linewidth,clip]{Fig7a.eps}
\end{minipage}
\hfill
\begin{minipage}[t]{0.475\linewidth}
\includegraphics[width=1.0\linewidth,clip]{Fig7b.eps}
\end{minipage}
\end{minipage}
\caption{Reflection coefficient~$R_0$ as function of the incident
 kinetic energy of singlet oxygen molecules, as calculated by the SH
 method with a $(a)$ minimal coupling; $(b)$ maximal coupling. The
 results both within the frozen substrate approximation (dashed lines)
 and for a three-dimensional surface oscillator (squares and triangles) are
 shown.  In the respective insets, the relative triplet
 yield~$R_T/R_0$ for both calculations is plotted as well. }
\label{SurfOsci}
\end{figure*}

To further substantiate our findings, we also inspected the role of
the surface mobility, which has not been accounted for in the hitherto
presented calculations. As mentioned before, the employed
potential-energy surfaces~\cite{Behler:2007p178} solely depend upon
the degrees of freedom of the oxygen but not on the ones of the
aluminum atoms. In this so called \textit{frozen substrate
 approximation} no energy dissipation into the bulk is possible, so
that our description of the dynamics close to the surface, where the
adsorption energy of the oxygen molecule is freed, is certainly
defective. For the actual triplet dissociation probability at low
incident energies, which is basically determined by the barriers
present in the entrance channel~(see Sec.~\ref{StickCoeff2}), the dynamics of
the bulk atoms can generally be neglected. Still, the frozen substrate
approximation might affect the reflection coefficients for the singlet
oxygen, in the dynamics of which backscattering processes close to the
surface play an important role.

To overcome the frozen substrate approximation we performed
simulations with an additional \textit{surface oscillator}~\cite{Bach:2001p341,Bach:2003p342,Carbogno:2007p338},
which can account for the possible energy transfer between molecule
and bulk in hindsight, i.e.,~on the basis of potential-energy surfaces
derived within the frozen substrate approximation. For this, the
aluminum slab is permitted to oscillate as a whole in a harmonic
potential, so that the actual molecule-surface distance does then not
solely depend on the cartesian center-of-mass coordinate~$z$ of the
oxygen anymore but also on the amplitude of the surface
oscillator~$\tilde{z}$. Thus both the diagonal and the non-diagonal
potential matrix elements that enter the surface hopping simulations
in Eq.~(\ref{eq:classical}) and (\ref{eq:coeff}) have to be formally
adjusted to reflect this additional dependence:
\begin{equation}
V_{ij}\left(\cdots,z,\cdots\right) \xrightarrow{\mbox{\footnotesize Surf. Osci.}} V_{ij}\left(\cdots,[z-\tilde{z}],\cdots\right) .
\label{SurfOsciAdj}
\end{equation}
Furthermore, the auxiliary harmonic potential of the slab, which does not depend on the electronic state, is added to each of the diagonal elements of the potential matrix
\begin{equation}
V_{ii}^{7D}(\cdots) = V_{ii}(\cdots) + \frac{\hbar \omega}{2}\tilde{z}^2,
\label{SurfOsciAdj2}
\end{equation}
whereas no additional coupling term is added to the non-diagonal
elements. Analogously, this model can be extended to also allow for an
oscillation of the slab with respect to the lateral surface
coordinates, as done in all the simulations presented below. Although
such a surface oscillator does not yield any additional information
about the actual motion of the individual aluminum atoms, it
effectively incorporates the energy transfer effects in between the
slab and the molecule into the simulations. For the frequency~$\omega$
of the harmonic potential in Eq.~(\ref{SurfOsciAdj2}) a value of
$70$\% of the experimental Debye frequency~($\approx 34$~meV) of the
bulk~\cite{AshcroftMermin} has been used, since previous studies
showed~\cite{Hand:1990p52,Gro:1993p915,Gro:1994p914} that this
fraction is best suited to model the energy dissipation. For the exact
same reason three times the mass of a single aluminum atom is assigned
to the surface oscillator to account for finite size
effects~\cite{Bach:2001p341,Bach:2003p342}. In the herein presented
calculations, a kinetic energy that corresponds to a temperature of
$330$~K has been assigned to the surface vibration in the initial
condition. However, no significant dependence on this parameter has
been found in the temperature regime at which the experiments are
typically performed~\cite{Brune:1992p222, Osterlund:1997p151,
 SCHMID:2001p208, Weie:2003p98}.

Not too surprisingly, the sticking coefficient for incident triplet
molecules is hardly affected by the surface oscillator at all due to
the fact that this observable is principally determined by the
dynamics in the entrance channel. As a matter of fact, absolute changes of
$\pm 3$\% due to the inclusion of the surface oscillator have been
found in the simulations at most~\cite{CarbognoPHD} and are thus not
further detailed in this work. In the high energy scattering regime discussed in
Sec.~\ref{HiNRG}, however, the already minute reflection coefficients
are even further diminished due to the surface mobility.

In contrast to this, the surface oscillator affects the reflection
coefficient for singlet molecules in a much more peculiar fashion that
sheds light on the underlying transition mechanisms: In the case of a
maximal coupling the effects are again minute, as shown in
Fig.~\ref{SurfOsci}: At large incident velocities, the 
surface mobility favors dissociation as intuitively expected, since
the motion of the aluminum atoms offers a dissipation channel for the
incident energy. At low velocities, the inclusion of the surface
oscillator leads to a slight increase in the reflectivity. Already
when climbing the barriers the molecules lose a small portion of
their kinetic energy to the surface oscillator due to Newton's second
law. In certain cases, this minute energy
transfer is already sufficient to hinder the surmounting of the
respective barriers. However, the net effect is again minute, since
these processes occur far away from the surface.

We find a more pronounced influence of the surface mobility in the
surface hopping calculations with a minimal coupling, as also shown in
Fig.~\ref{SurfOsci}. At large incident energies the reflection
probability drops to zero. The incident molecules become trapped close to the
surface due to the energy loss to the surface
oscillator when hitting the aluminum, and thus eventually
dissociate. The exact same mechanism is active at low incident
energies, as also discussed before. However, the energy dissipation
through the bulk slows down the oxygen even more, so that a transition
to the triplet state becomes more and more probable in spite of the
minute coupling value. Accordingly, we find an increase in both the
total and the relative yield when the mobility of the surface atoms is
accounted for.

\section{Conclusion}
\label{CONC}
In this work we extended previous MD studies~\cite{Behler:2005p19} of
the dissociative adsorption dynamics of oxygen on the aluminum
surface by including multiple PESs associated to different spin-configurations
in the dynamics by means of Tully's \textit{Fewest Switches}
algorithm. Apart from reproducing the experimental sticking coefficient at
normal incidence, this approach additionally allows to study the
characteristics of the underlying nonadiabatic transitions in
detail. This knowledge allows to propose experiments that are able to
unambiguously identify the occurrence of such spin transitions. Whereas the
detection of the triplet-to-singlet conversion in high energy
scattering experiments is highly unlikely, a non-zero probability for
the singlet-to-triplet conversion in scattering experiments with a
beam of singlet oxygen molecules has been found, even in the limit of
an unrealistically small coupling. If we additionally include the motion
of the substrate atoms in our model, the determined yields are either
not altered significantly or even enhanced. Depending on the coupling
strength, two different mechanisms, which result in different kinetic
energy distributions in the reflection channel, have been found in the
singlet-to-triplet conversion dynamics. An experimental determination
of these distributions does thus allow (a) to prove the
nonadiabaticity of this reaction and (b) to draw conclusions on the
character of the ongoing transitions.

\section*{Acknowledgments}
Financial grants from the \textit{Deutsche Forschungsgemeinschaft} within the 
projects RE 1509/7-1 and GR 1503/17-1 is gratefully acknowledged. 
JB is grateful for financial support by the FCI and the DFG (BE 3264/3-1).
Additional computational resources were provided by the D-GRID~(bwGRID) project.

\printfigures
\printtables

\end{document}